\documentstyle[preprint,aps,axodraw]{revtex}

\newcommand{\beq}{\begin{equation}}
\newcommand{\eeq}{\end{equation}}
\newcommand{\beqa}{\begin{eqnarray}}
\newcommand{\eeqa}{\end{eqnarray}}
\newcommand{\snu}{\tilde \nu}

\newcommand{\dmsnu}{{\mbox{$\Delta m_{\tilde \nu}$}}}

\def\npb#1{{\sl Nucl.\ Phys.}\ {\bf B#1}}
\def\plb#1{{\sl Phys.\ Lett.}\ {\bf B#1}}
\def\prd#1{{\sl Phys.\ Rev.}\ {\bf D#1}}
\def\prl#1{{\sl Phys.\ Rev.\ Lett.} {\bf #1}}

\def\prep#1{{\sl Phys.\ Rep.}\ {\bf #1}}

\def\ifmath#1{\relax\ifmmode #1\else $#1$\fi}
\def\half{\ifmath{{\textstyle{1 \over 2}}}}

\begin{document}

\draft{\tighten

\preprint{
\vbox{
      \hbox{SLAC-PUB-7423}
      \hbox{SCIPP 97/02}
      \hbox{hep-ph/9702421}}}

\bigskip
\bigskip

\title{Sneutrino Mixing Phenomena}
\author{Yuval Grossman\,$^a$ and  Howard E. Haber\,$^b$}
\address{ \vbox{\vskip 0.truecm}
$^a$Stanford Linear Accelerator Center \\
        Stanford University, Stanford, CA 94309 \\
\vbox{\vskip 0.truecm}
  $^b$Santa Cruz Institute for Particle Physics \\
University of California, Santa Cruz, CA 95064}

\maketitle

\begin{abstract}%
In any model with nonzero Majorana neutrino masses, the sneutrino and
antisneutrino of the supersymmetric extended theory mix.  We outline the
conditions under which sneutrino-antisneutrino mixing is
experimentally observable.
The mass-splitting of the sneutrino mass eigenstates and sneutrino
oscillation phenomena are considered.
\end{abstract}

}\newpage


\section{Introduction}
In the Standard Model, neutrinos are exactly massless \cite{revnu}.
However, a number of experimental hints suggest that neutrinos may have
a small mass.
The solar neutrino
puzzle can be solved by invoking the MSW mechanism, with the neutrino
squared-mass difference of 
$\Delta m^2\simeq 6\times 10^{-6}\,$eV$^2$ \cite{hata}.
The atmospheric neutrino puzzle could be explained if
$\Delta m^2\simeq 10^{-2}\,$eV$^2$ \cite{barish}.
The LSND experiment has reported a signal
that, if interpreted as neutrino oscillations, implies
$\Delta m^2\sim {\cal O} (1\,{\mbox{\rm eV}}^2)$ \cite{LSND}.
To accommodate this data, the Standard
Model must be extended; the simplest models simply add
Majorana neutrino mass terms that violate lepton number ($L$) by two
units.

One must also extend the Standard Model in order to accommodate light
Higgs bosons in a more fundamental unified theory that incorporates
gravity. Models of low-energy supersymmetry \cite{revsusy} are
attractive candidates for the theory of TeV scale physics.
However, in the minimal supersymmetric extension of the Standard Model
(MSSM), neutrinos are also exactly massless.  
In this paper, we wish to consider a
supersymmetric extension of an extended Standard Model that contains Majorana
neutrino masses.  In such models, the lepton number violation can
generate interesting phenomena in the sector of supersymmetric leptons.
The effect of $\Delta L=2$ operators is to introduce a mass 
splitting and mixing into the sneutrino--antisneutrino system 
(this observation was also made recently in Ref. \cite{HKK}).
The sneutrino and antisneutrino will then no
longer be mass eigenstates.

This phenomena is analogous to
the effect of a small $\Delta B=2$ perturbation
to the leading $\Delta B=0$ mass term in the $B$-system \cite{revB}.
This results in a mass splitting between the
heavy and light $B^0$ (which are no longer pure $B^0$ and $\bar{B^0}$
states).  The very small mass splitting, ${\Delta m_B /m_B} = 7 \times
10^{-14}$ \cite{pdg}, can be measured by observing flavor oscillations.
The flavor is tagged in $B$-decays
by the final state lepton charge.
Since $x_d \equiv \Delta m_B /\Gamma_B \approx 0.7$ \cite{pdg}, there is time for
the flavor to oscillate before the meson decays.  Then the time-integrated same
sign dilepton signal is used to determine $\Delta m_B$.

The sneutrino system can exhibit similar behavior. The lepton number is tagged
in sneutrino decay using the charge of the outgoing lepton.  The relevant scale
is the sneutrino width (as emphasized in the context of lepton flavor
oscillation in Ref.~\cite{ACFH}).  If the sneutrino mass splitting is
large, namely
\beq \label{xsnudef}
x_{\snu} \equiv {\dmsnu \over \Gamma_{\snu}} \gtrsim 1,
\eeq
and the sneutrino branching ratio into a charged lepton is significant,
then a measurable same sign dilepton signal is expected.

The neutrino mass and the
sneutrino mass splitting are related as a consequence of the lepton number
violating interactions and supersymmetry breaking.
Thus, we can use upper bounds (or indications)
of neutrino masses to set bounds on the sneutrino mass splitting.
At present, neutrino mass bounds obtained from direct
laboratory measurements imply \cite{pdg}:
$m_{\nu_e}\lesssim 10\,$eV,
$m_{\nu_\mu}\leq0.17\,$MeV, and
$m_{\nu_\tau}\leq24\,$MeV.
Cosmological constraints
require stable neutrinos to be lighter than about $100\,$eV.
For example, models of mixed 
dark matter require a neutrino mass of order
$10\,$eV \cite{Primack}.  For unstable neutrinos, the mass
limits are more complex and model-dependent \cite{HaNi}.  In this paper we
will consider the consequences of two cases: (i) $\nu_\tau$ with a mass near 
its present laboratory upper limit, and (ii) light neutrinos of mass less 
than 100~eV.

Some model-independent relations among the neutrino and sneutrino
$\Delta L=2$ masses (and other $\Delta L=2$ phenomena) have been
derived in Ref. \cite{HKK}.  However, in order to derive specific results, it
is useful to exhibit an explicit model of lepton number violation.  In
the following, we concentrate on the see-saw model for neutrino masses
\cite{revnu}, as it exhibits all the interesting features.  We compute the 
sneutrino mass splitting in this model and discuss its 
implications for sneutrino phenomenology at $e^+e^-$ colliders.
(We also briefly mention some consequences of lepton number violation 
arising from R-parity nonconservation.)  A more complete presentation
will be given in Ref. \cite{followup}.



\section{The Supersymmetric See-Saw Model}
Consider an extension of the MSSM where
one adds a right-handed neutrino
superfield, $\hat N$, with a bare mass $M \gg m_Z$.
For simplicity we consider a one generation model
({\it i.e.}, we ignore lepton flavor mixing) and assume CP conservation.  We
employ the most general R-parity conserving
renormalizable superpotential and attendant
soft-supersymmetry breaking terms.  For this work, the relevant terms
in the superpotential are (following the notation of
Ref.~\cite{habertasi})
\beq
W=\epsilon_{ij}\left[\lambda \hat H_2^i \hat L^j \hat N - \mu \hat H_1^i
\hat H_2^j\right]+ \half M\hat N \hat N\,.
\eeq
The $D$-terms are the same as in the MSSM.
The relevant terms in the soft-supersymmetry-breaking scalar potential
are:
\beq
V_{\rm soft}= m_{\tilde L}^2 \snu^*\snu + m_{\tilde N}^2 \tilde N^*
\tilde N  + 
(\lambda A_\nu H_2^2\snu\tilde N^* + MB_N \tilde N\tilde N + {\rm h.c.}) \,.
\eeq
When the neutral Higgs field vacuum
expectation values are generated [$\langle H_i^i\rangle=v_i/\sqrt{2}$, with
$\tan\beta\equiv v_2/v_1$ and $v_1^2+v_2^2=(246~{\rm GeV})^2$], one finds that
the light neutrino mass is given by the usual one generation see-saw result,
$m_\nu \simeq m_D^2/M$,
where $m_D\equiv\lambda v_2$ and we drop terms higher order in $m_D/M$.

The
sneutrino masses are obtained by diagonalizing a $4\times 4$ squared-mass
matrix.  Here, it is convenient to define: $\snu=(\snu_1+i\snu_2)/\sqrt{2}$ and
$\tilde N=(\tilde N_1+i\tilde N_2)/\sqrt{2}$.  Then,
the squared-sneutrino
mass matrix (${\cal M}^2$) separates into CP-even and CP-odd blocks:
\beq
{\cal M}^2=\half \pmatrix{\phi_1&\phi_2\cr}
\pmatrix {{\cal M}^2_+ & 0 \cr 0 & {\cal M}^2_-
\cr}\pmatrix{\phi_1\cr
\phi_2 \cr}\,,
\eeq
where $\phi_i\equiv \pmatrix{\snu_i&\tilde N_i\cr}$ and
\beq
{\cal M}^2_\pm = \pmatrix{
m_{\tilde L}^2 + \half m_Z^2 \cos 2\beta + m_D^2 &
m_D[A_\nu-\mu \cot\beta\pm M] \cr
m_D[A_\nu-\mu \cot\beta\pm M] & M^2 + m_D^2 + m_{\tilde N}^2 \pm 2 B_N M  
\cr}\,.
\eeq
In the following derivation we assume that $M$ is the largest
mass parameter. Then, to first order in $1/M$,
the two light sneutrino
eigenstates are $\snu_1$ and $\snu_2$, with corresponding squared masses:
\beq
m^2_{\snu_{1,2}} = m_{\tilde L}^2+
\half m_Z^2 \cos 2\beta\mp\half\Delta m^2_{\snu}\,,
\eeq
where the squared mass difference $\Delta m^2_{\snu}\equiv
m^2_{\snu_2}-m^2_{\snu_1}$ is of order $1/M$. Thus, in the large $M$ limit, we
recover the two degenerate sneutrino states of the MSSM,
usually chosen to be $\snu$ and $\bar{\snu}$.  
For finite $M$, these two
states mix with a $45^\circ$ mixing angle, since the two
light sneutrino mass eigenstates
must also be eigenstates of CP.  The sneutrino mass splitting is easily
computed using $\Delta m^2_{\snu}=2m_{\snu}\Delta m_{\snu}$, where
$m_{\snu}\equiv\half(m_{\snu_1}+m_{\snu_2})$ is the average of the light
sneutrino masses. We find that the ratio of the light
sneutrino mass difference relative to the light {\it neutrino} mass 
is given by (to leading order in $1/M$)
\beq \label{ratio}
r_\nu \equiv {\Delta m_{\snu} \over m_\nu} \simeq
{2 (A_\nu-\mu \cot\beta-B_N) \over m_{\snu}}\,.
\eeq

The magnitude of $r_\nu$ depends on various supersymmetric
parameters.
Naturalness constrains supersymmetric mass parameters
associated with particles with non-trivial electroweak quantum numbers
to be roughly of order $m_Z$ \cite{natural}.
Thus, we assume that
$\mu$, $A_\nu$, and $m_{\tilde L}$ are all of order the electroweak
scale.  The parameters $M$, $m_{\tilde N}$,
and $B_N$ are fundamentally different since
they are associated with the SU(2)$\times$U(1) singlet superfield
$\hat N$.  In particular, $M\gg m_Z$, since this drives the see-saw
mechanism.  Since $M$ is a supersymmetry-conserving parameter,
the see-saw hierarchy is technically natural.
The parameters $m_{\tilde N}$ and $B_N$ are
soft-supersymmetry-breaking parameters; their order of magnitude is less clear.
Since $\hat
N$ is an electroweak gauge group singlet superfield,
supersymmetry-breaking terms associated with it need
not be tied to the scale of electroweak symmetry breaking.
Thus, it is possible that $m_{\tilde N}$ and $B_N$ are much larger than $m_Z$.
Since $B_N$ enters directly into the formula for the
light sneutrino mass splitting [eq.~(\ref{ratio})],
its value is critical for sneutrino phenomenology.
If $B_N\sim{\cal O}(m_Z)$, then $r_\nu\sim {\cal O}(1)$, 
which implies that the sneutrino mass splitting is of order the {\it
neutrino} mass.  However, if $B_N\gg m_Z$, then the sneutrino mass splitting is
significantly enhanced.

We have also considered other possible models of lepton number
violation \cite{followup}.  
For example, in models of R-parity violation (but with no
right handed neutrino), a sneutrino
mass splitting is also generated whose magnitude
is of order the corresponding neutrino mass.  Thus, in models where
R-parity violation is the {\it only} source of lepton number violation,
$r_\nu\simeq{\cal O}(1)$, and no enhancement of the sneutrino mass
splitting is possible.

\section{Loop Effects}
In the previous section, all formulae given involved tree-level
parameters.
However, in some cases, one-loop effects can substantially modify
eq.~(\ref{ratio}). In general, the existence of a sneutrino mass splitting
generates a one-loop contribution to the neutrino mass.
Note that this effect is generic, and is independent of the mechanism that
generates the sneutrino mass splitting.
Similarly, the existence of a Majorana
neutrino mass generates a one-loop contribution
to the sneutrino mass splitting.  However, the latter effect can be safely
neglected.  Any one-loop contribution to the sneutrino mass splitting must
be roughly $\Delta m_{\snu}^{(1)}\sim(g^2/16\pi^2)m_\nu$; thus
the tree-level result $r_\nu\gtrsim {\cal O}(1)$ cannot be 
significantly modified.
In contrast, the one-loop correction to the neutrino mass is potentially
significant, and may dominate the tree-level mass 
[$m_\nu^{(0)}\simeq m_D^2/M$].
We have computed exactly the one-loop contribution to the neutrino mass
[$m_\nu^{(1)}$] from neutralino/sneutrino loops shown 
in Fig.~1. In the limit of
$m_\nu,\Delta m_{\snu}\ll m_{\snu}$,
the formulae simplify, and we find
\beq \label{loopmass}
m_\nu^{(1)} = {g^2\Delta m_{\snu} \over 32 \pi^2 \cos^2 \theta_W}
\sum_j \,f(y_j) |Z_{jZ}|^2\,, \qquad 
f(y_j) = {\sqrt{y_j}\left[y_j-1-\ln(y_j)\right]\over (1-y_j)^2}\,,
\eeq
where $y_j \equiv {m_{\snu}^2/m_{\tilde\chi^0_j}^2}$ and
$Z_{jZ}\equiv Z_{j2}\cos\theta_W-Z_{j1}\sin\theta_W$ is the
neutralino mixing
matrix element that projects out the $\tilde Z$ eigenstate from the $j$th
neutralino. One can check that $f(y_j)<0.566$, and
for typical values of $y_j$ between 0.1 and 10, $f(y_j) > 0.25$.
Since $Z$ is a unitary
matrix, we find $m_\nu^{(1)}\approx  10^{-3} m_\nu^{(0)} r_\nu^{(0)}$, where
$r_\nu^{(0)}$ is the tree-level ratio computed in eq.~(\ref{ratio}). If
$r_\nu^{(0)}\gtrsim 10^3$, then the one-loop contribution to the neutrino mass
cannot be neglected.  Moreover, $r_\nu$ cannot be arbitrarily large
without unnatural fine-tuning.  Writing the neutrino mass as
$m_\nu= m_\nu^{(0)}+m_\nu^{(1)}$, and assuming no unnatural
cancellation between the two terms, we conclude that
\beq
r_\nu \equiv {\Delta m_{\snu} \over m_\nu} \lesssim 2\times 10^{3}.
\eeq


\section{Phenomenological Consequences}
Based on the analysis presented above, we take
$1 \lesssim r_\nu \lesssim 10^3$.  If $r_\nu$ is near its maximum, and if there
exists a neutrino mass in the MeV range, then the corresponding sneutrino mass
difference is in the GeV range.  Such a large mass splitting can be observed
directly in the laboratory.  For example, in $e^+e^-$ annihilation,
third generation sneutrinos
are produced via $Z$-exchange. Since the two sneutrino mass
eigenstates are CP-even and CP-odd respectively, 
sneutrino pair production
occurs only via $e^+e^-\to \snu_1\snu_2$.
In particular, the pair production processes 
$e^+e^-\to\snu_i\snu_i$ (for $i=1,2$) are forbidden.
If the low-energy supersymmetric model incorporates some R-parity
violation, then
sneutrinos can be produced in $e^+e^-$ via an $s$-channel resonance
\cite{EFP,BGH}.  Then, for a sneutrino mass difference in the GeV range,
two sneutrino resonant peaks could be distinguished.

A smaller sneutrino mass splitting can be probed in $e^+e^-$ annihilation
using the same sign dilepton
signal if $x_{\snu} \gtrsim 1$.  Here we must rely on sneutrino oscillations.
Assume that the sneutrino decays with significant branching ratio via
chargino exchange:  $\snu \to \ell^\pm + X$.  Since this decay
conserves lepton number, the lepton number
of the decaying sneutrino is tagged by the lepton charge.  Then in
$e^+e^-\to\snu_1\snu_2$, the probability of a same sign dilepton signal
is
\beq
P(\ell^+\ell^+) + P(\ell^-\ell^-)= \chi_{\snu}
[BR(\snu \to \ell^\pm +X)]^2\,,
\eeq
where $\chi_{\snu} \equiv x_{\snu}^2/[2(1+x_{\snu}^2)]$ is 
the integrated oscillation probability, which
arises in the same way as the corresponding quantity that appears in the
analysis of $B$ meson oscillations \cite{revB}.
We have considered the constraints on the supersymmetric model imposed
by the requirements that 
$x_{\snu}\sim{\cal O}(1)$ and BR$(\snu\to\ell^\pm+X)\sim 0.5$.
We examined two cases depending on
whether the dominant $\snu$ decays involve two-body or three-body final states.

If the dominant sneutrino decay involves two-body final states, then we must
assume that $m_{\tilde\chi_1^0}< m_{\tilde\chi^+} <m_{\snu}$.
Then, the width of the two leading sneutrino decay channels
are (neglecting the mass
of the charged lepton) \cite{BGH,gunhab}
\beqa \label{gammas}
\Gamma(\snu \to \tilde\chi_j^0 \nu)& =&
{g^2|Z_{iZ}|^2 m_{\snu}\over 32 \pi \cos^2 \theta_W}
B(m_{\tilde\chi_j^0}^2/m_{\snu}^2)\,, \\
\Gamma(\snu \to \tilde\chi^+ \ell^-) &=& {g^2|V_{11}|^2 m_{\snu}\over 16 \pi}
B(m_{\tilde\chi^+}^2/m_{\snu}^2)\,, \nonumber
\eeqa
where $B(x)=(1-x)^2$,
$V_{11}$ is one of the mixing matrix elements in the chargino sector,
and $Z_{jZ}$ is the neutralino mixing matrix element
defined below eq.~(\ref{loopmass}). For example,
for $m_{\snu} \sim {\cal O}(m_Z)$ we find
$\Gamma(\snu \to \chi_j^0 \nu) \approx {\cal O}(|Z_{jZ}|^2
B(m_{\tilde\chi_j^0}^2/m_{\snu}^2)\times 1\,
{\mbox{\rm GeV}})$ and
$\Gamma(\snu \to \chi^+ \ell) \approx  {\cal O}(|V_{11}|^2
B(m_{\tilde\chi^+}^2/m_{\snu}^2)\times 1 \,{\mbox{\rm GeV}})$.
We require that the sneutrino and
chargino are sufficiently separated in mass, so
that the emitted charged lepton will not be too soft
and can be identified experimentally.  This implies that
$B\gtrsim 10^{-2}$ in eq.~(\ref{gammas}).  Thus,
for the third generation sneutrino, a significant same-sign dilepton
signal can be generated with $m_{\nu_\tau}= 10\,$MeV,
even if $r_\nu \sim 1$
and the light chargino/neutralino mixing angles are of ${\cal O}(1)$.
If the lightest chargino and two lightest neutralinos
are Higgsino-like, then the mixing angle factors in eq.~(\ref{gammas}) are
suppressed.  For $|\mu|\sim m_Z$ and gaugino mass parameters not larger
than $1\,$TeV, the square of the light chargino/neutralino mixing angles
must be of ${\cal O}(10^{-2})$ or larger.  Thus, if
$r_\nu$ is near its maximum value ($r_\nu \sim 10^3$), then one
can achieve $x_{\snu} \sim 1$ for {\it neutrino} masses as low as about
$100\,$eV.

If no open two-body decay channel exists, then we must consider the
possible sneutrino decays into three-body final states.  In this case
we require that $m_{\snu}<m_{\tilde\chi_1^0},m_{\tilde\chi^+}$.  Again, we
assume that there exists a significant chargino-mediated
decay rate with charged leptons in the final state.
The latter occurs in models in which the $\tilde\tau_R$
is lighter than the sneutrino.  In this case, the rate for chargino-mediated
three-body decay $\snu_\ell\to\tilde\tau_R\nu_\tau\ell$ can be
significant.  The $\tilde\tau_R$ with $m_{\tilde\tau_R}<m_{\snu}$
can occur in
radiative electroweak breaking models of low-energy supersymmetry if
$\tan\beta$ is large.  However, in the context of the MSSM, such a
scenario would require that $\tilde\tau_R$ is the lightest
supersymmetric particle (LSP), a possibility strongly disfavored
by astrophysical bounds on the abundance of stable heavy charged particles
\cite{staulsp}. Thus, we go beyond the usual MSSM assumptions and assume
that the $\tilde\tau_R$ decays.  This can occur in gauge-mediated
supersymmetry breaking models \cite{DDRT} where
$\tilde\tau_R\to\tau +\tilde g_{3/2}$, or in R-parity violating
models where $\tilde\tau_R \to\tau\nu$.
Here, we have assumed that intergenerational lepton
mixing is small;
otherwise the $\Delta L=2$ sneutrino mixing effect is diluted.

We have computed the chargino and neutralino-mediated three-body
decays of $\snu_\ell$.
In the analysis presented here, we have not considered the case of
$\ell=\tau$, which involves a more complex final
state decay chain containing two $\tau$-leptons.
For simplicity, we present analytic formulae in the limit where the
mediating chargino and neutralinos are much heavier than the
$\tilde\tau_R^\pm$.  In addition, we assume that the
lightest neutralino is dominated by its bino component.
We have checked that our conclusions
do not depend strongly on these approximations.
Then, the rates for the chargino and neutralino-mediated
sneutrino decays are:
\beqa
\Gamma(\snu_\mu \to \tilde \tau \nu_\tau \mu) &=&
{g^4 m_{\snu}^3 m_\tau^2 \tan^2\beta \over
3 \times 2^9  \pi^3 (m_W^2 \sin 2\beta - M_2 \mu)^2}
\,f_{\tilde \chi^+}(m_{\tilde \tau}^2/m_{\snu}^2)\,, \nonumber \\
\Gamma(\snu_\mu \to \tilde \tau \nu_\mu \tau)&=&
{g'^4 m_{\snu}^5\over
3 \times 2^{11}  \pi^3 M_1^4}\, f_{\tilde
\chi^0}(m_{\tilde \tau}^2/m_{\snu}^2) \,,
\eeqa
where the $M_i$ are gaugino mass parameters,
$f_{\tilde \chi^+}(x) = (1-x)(1+10 x+x^2) + 6x(1+x) \ln x$ and
$f_{\tilde \chi^0}(x) = 1-8 x + 8 x^3 + x^4 + 12 x^2 \ln x$.
Assuming $\tan\beta\gtrsim 20$ (since $\tilde\tau_R$ is light as
mentioned above), and taking typical values for the other supersymmetric
parameters, we find that
$\Gamma(\snu_\mu \to \tilde \tau \nu_\tau \mu) \sim
\Gamma(\snu_\mu \to \tilde \tau \nu_\mu \tau)\sim {\cal O}
(1\,{\mbox{\rm eV}})$.
Thus, for $r_\nu \sim 1$ [$10^3$], a significant like-sign dilepton signal
could be observed for light neutrino masses as low as $1\,$eV [$10^{-3}\,$eV].


\section{Conclusions}
Non-zero Majorana neutrino masses imply the existence of
$\Delta L=2$ phenomena.  In low-energy supersymmetric models, such
phenomena also leads to sneutrino-antisneutrino mixing with
the corresponding mass eigenstates split in mass.
The sneutrino mass splitting is generally of the same order as the light
neutrino mass, although an enhancement of up to three
orders of magnitude is conceivable.
If the mass of the $\nu_\tau$ is near its present experimental bound,
then it may be possible to directly observe the sneutrino mass splitting
in the laboratory.  Even if neutrino masses are small (of order $1\,$eV),
some supersymmetric models yield an observable sneutrino oscillation
signal at $e^+e^-$ colliders.  Remarkably, model parameters exists
where sneutrino mixing phenomena are detectable for {\it neutrino}
masses as low as $m_\nu \sim 10^{-3}$\,eV
(a mass suggested by the solar neutrino
anomaly).  Thus, sneutrino mixing and oscillations could provide a novel
opportunity to probe lepton-number violating phenomena in the laboratory.

\acknowledgments
This project was initiated after a number of conversations with Jens
Erler.  
We thank him for his probing questions and insights.
We are also grateful to Jonathan Feng, Scott Thomas and Jim Wells for helpful
discussions. This work was supported by the U.S. Department of Energy
under contract numbers
YG is supported by the U.S. Department of Energy under
contract 
DE-AC03-76SF00515.
HEH is supported in part by the U.S. Department of Energy
under contract 
DE-FG03-92ER40689.

{\tighten

}
\begin{figure}
\begin{center}
\begin{picture}(200,80)(0,0)
\CArc(90,20)(30,0,180)
\DashLine(60,20)(120,20){7}
\Line(60,20)(20,20)
\Line(120,20)(160,20)
\Vertex(60,20){2}
\Vertex(120,20){2}
\Text(30,10)[]{$\nu$}
\Text(150,10)[]{$\nu$}
\Text(90,60)[]{$\chi^0_j$}
\Text(90,10)[]{$\tilde \nu_{1,2}$}
\end{picture}
\end{center}
\caption[a]{One-loop contribution to the neutrino mass due to
sneutrino mass splitting.}
\end{figure}
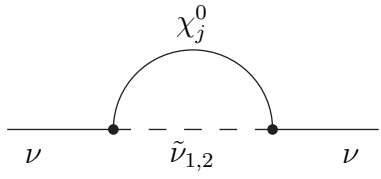

\end{document}